\documentclass[sigconf,natbib=true]{acmart}

\usepackage{graphicx}
\usepackage{booktabs}
\usepackage{bbm}


\copyrightyear{2026}
\acmYear{2026}
\setcopyright{cc}
\setcctype{by}
\acmConference[CIKM '26]{Proceedings of the 35th ACM International Conference on Information and Knowledge Management}{November 07--11, 2026}{Rome, Italy}
\acmBooktitle{Proceedings of the 35th ACM International Conference on Information and Knowledge Management (CIKM '26), November 07--11, 2026, Rome, Italy}
\acmDOI{10.1145/3799682.3839919}
\acmISBN{979-8-4007-2539-5/2026/11}
\begin{document}

\title{SCAR: Semantic Continuity-Aware Retrieval for Efficient Context Expansion in RAG}

\author{Nathana\"el Langlois}
\email{nathanael.langlois@horizon-flow.com}
\affiliation{%
  \institution{Horizon Flow}
  \city{Paris}
  \country{France}
}

\begin{abstract}
Fixed-length chunking in Retrieval-Augmented Generation (RAG) often leads to boundary fragmentation, where critical evidence is split across segments, degrading retrieval recall. While static windowing and parent retrieval improve recall, they introduce significant token overhead. We propose \textbf{SCAR} (Semantic Continuity-Aware Retrieval), an adaptive retrieval policy that selectively expands neighboring chunks by weighing query-neighbor relevance against a structural continuity penalty. SCAR uses a relative expansion threshold tied to each retrieved chunk's own query-relevance, yielding an approximately scale-invariant decision rule that transfers across embedding models without recalibration. Across four diverse corpora (RFC, GDPR, a 10-K report, and a Merger agreement; $N=320$ queries; 160 boundary-fragmented), SCAR achieves 92.8\% recall on boundary-fragmented queries with only 7.84 chunks, a 22.9\% reduction compared to static windowing (10.16 chunks). Paired bootstrap tests (B=10{,}000) confirm the chunk reduction is highly significant ($p<0.0001$, Cohen's $d=-1.49$, large effect), with a small recall difference (Cohen's $d=-0.33$). The policy transfers across three embedding models (\texttt{text-embedding-3-large}, \texttt{BGE-large-en-v1.5}, \texttt{zembed-1}) using the same single hyperparameter setting, and downstream RAGAS evaluation on the 10-K corpus confirms SCAR preserves generation faithfulness while reducing context tokens by 27.1\%.
\end{abstract}


\maketitle

\section{Introduction}
Retrieval-Augmented Generation (RAG) has emerged as the standard for grounding Large Language Models (LLMs) in external knowledge. However, document chunking remains a fundamental bottleneck in the retrieval pipeline. Standard strategies, such as recursive character splitting, prioritize local structural integrity but create a rigid, static index. This leads to the boundary fragmentation problem, where critical evidence required for a specific query is split across adjacent segments. Because optimal chunking is query-dependent and constrained by the inherent structure of the source documents, even advanced paragraph-aware splitters cannot prevent logical context breaks at runtime.

\section{Problem Statement}

In technical, legal, financial, or other complex structured documents, critical evidence often spans across standard indexing boundaries. When a retrieval system returns only one fragment of a bifurcated logical unit (e.g., the header of a financial table without its corresponding rows, or the first half of a legal definition), the downstream LLM suffers from incomplete context. 

\textit{Boundary Fragmentation} occurs when the evidence $E$ required to answer query $Q$ is split such that $E = \{c_i, c_{i+1}\}$, but the retriever only identifies $c_i$ as relevant. While a dense retriever may occasionally capture both chunks in a Top-$k$ result, it offers no guarantee of logical continuity. Standard solutions like \textbf{Static Windowing} (always retrieving $c_{i \pm 1}$) or \textbf{Parent Retrieval} (retrieving the entire section) are token-inefficient. These methods often inflate the context window with irrelevant "noise," which can degrade LLM performance and increase inference costs.

\section{Related Work}

Existing approaches to boundary fragmentation cluster into two camps. \textbf{Indexing-time coherence} methods build semantically continuous units before retrieval (e.g., \textbf{TextTiling} \cite{10.5555/972684.972687} via lexical cohesion, \textbf{Max--Min semantic chunking} \cite{kiss2025maxmin} via sentence-embedding similarity, \textbf{Late Chunking} \cite{jina2024} via long-context contextualized embeddings, and \textbf{SentGraph} \cite{liang2026sentgraphhierarchicalsentencegraph} via hierarchical sentence graphs), but fix boundaries at indexing time and cannot adapt to query-specific informational density. \textbf{Retrieval-time expansion} methods complement this. Static heuristics (fixed sliding windows, parent--child retrieval) deterministically include neighbors and are token-inefficient. Iterative methods (\textbf{FLARE} \cite{jiang-etal-2023-active}, \textbf{Self-RAG} \cite{asai2024selfrag}, \textbf{DIVER} \cite{long2025divermultistageapproachreasoningintensive}) interleave retrieval with generation or reranking through repeated LLM calls. Utility-trained retrievers such as \textbf{SCARLet} \cite{xu2026trainingutilitybasedretrievershared} improve quality but require specialized data synthesis and training. SCAR fills the gap as a single-shot, training-free retrieval-time policy that adaptively expands only when query relevance and structural continuity jointly justify it, and it composes orthogonally with any indexing-time chunking strategy.

\section{Methodology: SCAR}
We introduce \textbf{SCAR} (Semantic Continuity-Aware Retrieval), an adaptive retrieval policy designed to selectively recover fragmented logical context at runtime. 

\subsection{Continuity Score}
For a retrieved chunk $c$ and a candidate neighbor $n$, we define a boundary penalty:
\begin{equation}
b_{c,n} = 1 - \cos(e_c, e_n)
\end{equation}
where $e_c, e_n$ are the chunk embeddings. The boundary penalty $b_{c,n}$ quantifies the semantic discontinuity between adjacent chunks; a higher cosine similarity indicates stronger continuity.

\subsection{Expansion Decision}
We define the expansion score for a neighbor $n$ of $c$:
\begin{equation}
S_{c,n} = \cos(e_q, e_n) - \lambda \, b_{c,n}
\end{equation}

where $e_q$ is the query embedding and $\lambda$ is a hyperparameter controlling the strength of the continuity penalty.

A neighbor $n$ is appended to the context if its expansion score exceeds a \emph{relative} threshold tied to the retrieved chunk's own query-relevance:
\begin{equation}
S_{c,n} > \gamma \cdot \cos(e_q, e_c)
\label{eq:expand}
\end{equation}
where $\gamma \in (0,1)$ is a ratio hyperparameter. Intuitively, we expand $n$ if, after the continuity penalty, it is at least $\gamma$ as query-relevant as the chunk it is adjacent to.

This formulation has two design properties. \textbf{Scale robustness:} the relevance terms on both sides of Eq.~\ref{eq:expand} scale with query-chunk similarity, so the decision is approximately robust to the absolute similarity scale of the embedding model (exactly so in the $\lambda{=}0$ limit; for small $\lambda$ the continuity penalty introduces only a minor scale-dependent correction), largely removing the need for per-model threshold recalibration, which we confirm empirically in Section~\ref{sec:hyper}. \textbf{Adaptive thresholding:} the expansion bar adapts per retrieved chunk, since a weakly-retrieved chunk (low $\cos(e_q, e_c)$) triggers a correspondingly lower threshold, which is desirable when the retriever is itself uncertain.

\section{Experimental Setup}
\label{sec:experimental_setup}

\subsection{Datasets and Preprocessing}
We evaluate SCAR on four diverse corpora selected for their complex structural dependencies: the TCP specification (RFC 9293), the GDPR regulation, a Microsoft 10-K annual report, and a corporate merger agreement (Table~\ref{tab:datasets}). To prepare the documents, we utilize a hierarchical chunking strategy with \textbf{Contextual Prepending}. We employ a recursive character splitter with a target chunk size of 600 characters and a 60-character overlap. 

For each logical segment, we extract the structural hierarchy (e.g., \textit{Section > Subsection}) and prepend this metadata to the text fragment before embedding. This ensures that each chunk carries latent structural information, providing a rigorous baseline for our adaptive expansion policy.

\begin{table}[h]
\centering
\caption{Statistics of evaluation corpora.}
\label{tab:datasets}
\begin{tabular}{llc}
\toprule
\textbf{Corpus} & \textbf{Domain} & \textbf{Total Chunks} \\
\midrule
TCP            & Technical   & 273 \\
GDPR           & Legal       & 797 \\
Microsoft 10-K & Financial   & 654 \\
Merger         & Contractual & 461 \\
\bottomrule
\end{tabular}
\end{table}

\subsection{Retrieval and Evaluation}
We employ the \texttt{text-embedding-3-large} model to generate 3072-dimensional embeddings for all chunks and queries. Retrieval is performed using $k$-NN search with $k=5$ and cosine similarity. 

To evaluate SCAR, we generated 80 queries per corpus ($N=320$) using an LLM-assisted pipeline, followed by expert manual verification and a three-rater inter-annotator-agreement study (described below). This set is balanced between:
\begin{itemize}
    \item \textbf{Atomic Queries (40 per corpus)}: Information is contained within a single chunk.
    \item \textbf{Boundary-Fragmented Queries (40 per corpus)}: Information spans two or more contiguous chunks, requiring logical expansion for a complete answer.
\end{itemize}

Each query is paired with a set of \textbf{gold chunk IDs} for ground-truth verification. Our primary metric is \textbf{Recall}, defined as the proportion of gold chunks successfully recovered within the dynamically expanded context window. We evaluate this alongside \textbf{Chunk Efficiency}, defined as Recall divided by the average number of unique chunks retrieved per query: higher is better, capturing recall achieved per unit of context volume.
Our code and evaluation datasets are available at \url{https://github.com/scarmethod/SCAR}

\subsubsection*{Annotation Quality and Inter-Annotator Agreement.}
To validate the LLM-generated gold labels, three annotators independently identified relevant chunks without access to the gold: one labeled a stratified sample of 79 queries ($\sim$20 per corpus, balanced across query types) and two labeled overlapping ${\sim}20$-query subsets of it. Per-chunk Cohen's $\kappa$ against the gold was 0.71 and 0.79 for the two careful annotators (both substantial) and 0.57 for a third rapid pass; 91\% of queries were rated well-formed. Adjudication produced three benchmark corrections (two gold-set edits, one query reformulation), applied in all reported results.
\subsection{Baselines}
We compare SCAR against:
\begin{itemize}
    \item \textbf{Baseline (Top-$k$)}: Retrieve only the top-$k$ chunks
    \item \textbf{Window ($\pm 1$)}: Always retrieve immediate neighbors
    \item \textbf{Window ($\pm 2$)}: Expand to $\pm 2$ neighbors
    \item \textbf{Parent Retrieval}: Retrieve all chunks from the same document section
    \item \textbf{Cross-Encoder Reranker}: Rerank the top-20 dense candidates with \texttt{bge-reranker-large} and keep the top-$n$
\end{itemize}

\subsection{Hyperparameters}
SCAR has two hyperparameters: the boundary-penalty weight $\lambda$ and the relative-threshold ratio $\gamma$ from Eq.~\ref{eq:expand}. We fix $\lambda = 0.1$ and $\gamma = 0.80$ across all datasets and embedding models. We evaluate SCAR at expansion radius $r \in \{1, 2\}$. We verify the stability of these choices through a leave-one-corpus-out grid search and confirm that $\gamma=0.80$ transfers across embedding models without recalibration (Section~\ref{sec:hyper}).

\section{Results and Discussion}

\subsection{Main Results: Boundary-Fragmented Queries}

Table~\ref{tab:main_results} evaluates SCAR against the static windowing baseline on queries where evidence is split across segments. SCAR ($\pm 1$) achieves 92.8\% average recall using 7.84 unique chunks on average. In comparison, the Static Window baseline requires 10.16 chunks: a 22.9\% reduction in context volume with a 3.9 percentage point recall gap. Against naive fixed-budget retrieval, SCAR's selectivity is also evident: a top-8 retriever reaches only 0.905 recall (2.3 points below SCAR), and matching SCAR's 0.928 needs top-10 (about 28\% more chunks). The savings reflect adaptive selection, not a smaller candidate set.

The gain is most pronounced on the Microsoft 10-K corpus (29.9\% chunk reduction, 7.47 vs 10.65, at 92.3\% recall vs Window's 99.2\%), where SCAR filters the redundant ``neighbor noise'' common in dense financial tables.

\begin{table}[h]
\centering
\caption{Recall and average chunk volume on boundary-fragmented queries ($N=160$, $k=5$, $\lambda=0.1$, $\gamma=0.80$). W = Window ($\pm 1$), S = SCAR ($\pm 1$); chunk counts are unique segments after deduplication.}
\label{tab:main_results}
\begin{tabular}{lcccc}
\toprule
& \multicolumn{2}{c}{\textbf{Recall}} & \multicolumn{2}{c}{\textbf{Chunks}} \\ \cmidrule(lr){2-3} \cmidrule(lr){4-5}
\textbf{Dataset} & \textbf{W} & \textbf{S} & \textbf{W} & \textbf{S} \\
\midrule
TCP              & 0.992          & 0.950          & 9.75           & 7.90           \\
GDPR             & 0.941          & 0.926          & 10.57          & 8.10           \\
Microsoft 10-K   & 0.992          & 0.923          & 10.65          & 7.47           \\
Merger           & 0.943          & 0.914          & 9.68           & 7.88           \\
\midrule
\textbf{Average} & \textbf{0.967} & \textbf{0.928} & \textbf{10.16} & \textbf{7.84}  \\
\bottomrule
\end{tabular}
\end{table}

\begin{figure}[h]
\centering
\includegraphics[width=\columnwidth]{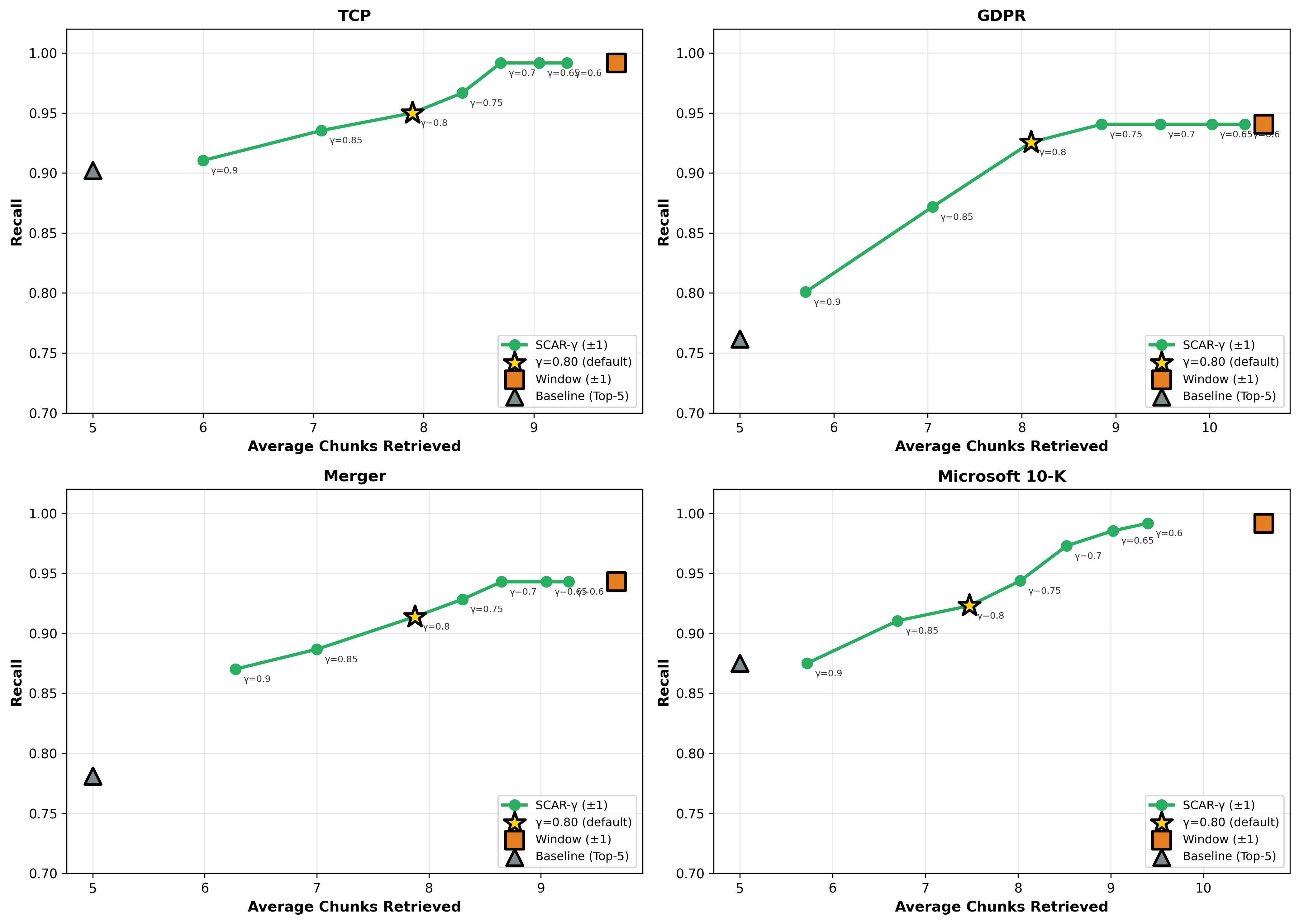}
\caption{Recall-chunk Pareto curves for SCAR-$\gamma$ ($\pm 1$) as $\gamma$ varies in $[0.60, 0.90]$ ($N=40$ boundary-fragmented queries per corpus). The gold star marks our $\gamma=0.80$ default, the orange square static Window ($\pm 1$), and the gray triangle the no-expansion Top-5 baseline. SCAR Pareto-dominates static windowing: at low $\gamma$ it matches Window's recall exactly in all four corpora while retrieving 1.0--1.7 fewer chunks, and $\gamma$ exposes a recall--cost dial that the static baselines lack.}
\Description{A two-by-two grid of line charts for the TCP, GDPR, Merger, and Microsoft 10-K corpora, each plotting recall from 0.70 to 1.00 against average chunks retrieved. A green curve traces SCAR as gamma decreases from 0.90 to 0.60, recall rising as more chunks are retrieved. In every panel the curve reaches the recall of the orange static-Window marker one to two chunks further left, and stays above the gray Top-5 baseline triangle fixed at five chunks. A gold star marks the gamma equals 0.80 default mid-curve.}
\label{fig:pareto}
\end{figure}

Figure~\ref{fig:pareto} shows Pareto curves demonstrating the recall-chunk tradeoff as the expansion threshold varies. SCAR's adaptive expansion consistently provides superior token efficiency at comparable recall to static windowing across the efficiency frontier. The gold star marks our chosen operating point under the relative-threshold formulation ($\gamma=0.80$).

\subsection{Method Comparison and Expansion Radius}

Table~\ref{tab:radius_analysis} compares all methods on the boundary-fragmented set. To assess whether SCAR's adaptive policy justifies larger expansion radii, we compare SCAR ($\pm 2$) against Window ($\pm 2$): while Window ($\pm 2$) blindly retrieves all neighbors (14.4 chunks average), SCAR ($\pm 2$) selectively expands only to semantically continuous chunks (9.2 vs 14.4), a 36\% chunk reduction at only 3.6 percentage points lower recall (0.949 vs 0.985), translating to roughly 50\% higher chunk efficiency.

\begin{table}[h]
\centering
\caption{Efficiency comparison on boundary-fragmented queries ($N=160$, $k=5$, $\lambda=0.1$, $\gamma=0.80$). The reranker keeps the top-8 of 20 reranked candidates, matching SCAR's chunk budget.}
\label{tab:radius_analysis}
\begin{tabular}{lccc}
\toprule
\textbf{Method} & \textbf{Recall} & \textbf{Chunks} & \textbf{Efficiency} \\
\midrule
Baseline (Top-$k$)               & 0.830 & 5.0   & 0.166 \\
\textbf{SCAR ($\pm 1$)}          & \textbf{0.928} & \textbf{7.84} & \textbf{0.118} \\
Window ($\pm 1$)                 & 0.967 & 10.16 & 0.095 \\
SCAR ($\pm 2$)                   & 0.949 & 9.2   & 0.103 \\
Window ($\pm 2$)                 & 0.985 & 14.4  & 0.069 \\
Reranker $\rightarrow$ top-8     & 0.854 & 8.0   & 0.107 \\
Parent                           & 0.968 & 50.2  & 0.019 \\
\bottomrule
\end{tabular}
\end{table}

SCAR dominates the efficiency frontier at both radii, offering recall-chunk tunability that static methods lack.

\textbf{Beyond expansion: cross-encoder reranking.} As a non-expansion alternative, we reranked the top-20 dense candidates with a widely-used cross-encoder (\texttt{bge-reranker-large}) and kept the top-8. Although $97.2\%$ of gold chunks lie within the candidate pool (Recall@20), the reranker reaches only $0.854$ recall at top-8 and $0.878$ at top-10 (Table~\ref{tab:radius_analysis}): above naive top-5 dense ($0.830$) but well short of SCAR's $0.928$ at the same budget. Even a high-quality point-wise scorer tends to demote \emph{continuation} chunks, whose standalone query relevance is lower than their keyword-bearing boundary partners. This suggests the gap likely stems from the ranking objective rather than candidate availability: adjacent-boundary recovery appears to benefit from structural-continuity reasoning beyond point-wise ranking alone.

SCAR's expansion is also query-dependent: on atomic queries ($N=160$), SCAR retrieves 7.6 chunks on average versus Window's 11.3, a 58.2\% reduction in \emph{unnecessary} expansion above the Top-5 baseline, providing a ``zero-regret'' profile when expansion is not required.

\subsection{Ablation Study}
To evaluate the individual contribution of the continuity penalty $b_{c,n}$, we compare the full SCAR policy ($\lambda=0.1$) against a \textit{Relevance-Only} expansion policy ($\lambda=0$). This ablation determines whether query relevance alone is sufficient for selective expansion or if the structural continuity signal is required to prune redundant neighbors.

As shown in Table~\ref{tab:ablation}, removing the continuity penalty increases chunk volume by up to 7.2\% with little recall change. The full SCAR policy consistently reduces overhead across all four corpora. This confirms that the continuity penalty acts as an effective semantic filter, suppressing irrelevant adjacent content. Finally, embedding raw chunk text \emph{without} the prepended structural metadata preserves SCAR's chunk-efficiency advantage (37.1\% reduction vs Window, versus 22.9\% with metadata), confirming the continuity penalty reflects content continuity rather than breadcrumb overlap.

\begin{table}[h]
\centering
\caption{Ablation Study: Impact of Continuity Penalty ($\lambda$) at $k=5, \gamma=0.80$, $N=160$ boundary-fragmented queries. Chunks represent the average unique segments retrieved.}
\label{tab:ablation}
\begin{tabular}{@{}lcccc@{}}
\toprule
& \multicolumn{2}{c}{\textbf{Recall}} & \multicolumn{2}{c}{\textbf{Avg. Chunks}} \\ \cmidrule(lr){2-3} \cmidrule(lr){4-5}
\textbf{Dataset} & \textbf{$\lambda=0$} & \textbf{$\lambda=0.1$} & \textbf{$\lambda=0$} & \textbf{$\lambda=0.1$} \\ \midrule
TCP              & 0.967              & 0.950              & 8.35              & 7.90              \\
GDPR             & 0.941              & 0.926              & 8.68              & 8.10              \\
Microsoft 10-K   & 0.944              & 0.923              & 7.95              & 7.47              \\
Merger           & 0.928              & 0.914              & 8.25              & 7.88              \\ \bottomrule
\end{tabular}
\end{table}

\subsection{Statistical Significance}
\label{sec:sig}

We test whether the observed chunk-volume and recall differences between SCAR ($\pm 1$) and Window ($\pm 1$) are statistically reliable. Because each query is evaluated under both methods on the same retrieval state, all comparisons are \emph{paired}.

For each query $q$ we record the per-query paired differences $\Delta_q^{\text{chunks}} = |S_q| - |W_q|$ and $\Delta_q^{\text{recall}} = \text{rec}_S(q) - \text{rec}_W(q)$. We compute: (i) a paired bootstrap test with $B=10{,}000$ resamples of the pair indices, yielding both a two-sided $p$-value and a 95\% confidence interval; (ii) a Wilcoxon signed-rank test as a non-parametric complement; and (iii) Cohen's $d$ on the paired differences as a standardized effect size.

On the boundary-fragmented set ($N=160$), the chunk reduction is highly significant: mean difference $-2.32$ chunks, 95\% bootstrap CI $[-2.57, -2.09]$, $p_\text{boot}<10^{-4}$, $p_\text{wilcox}<10^{-4}$, Cohen's $d=-1.49$ (large effect). Per-corpus tests all yield $p<10^{-4}$ on chunks. The recall difference is smaller: mean $-0.039$, 95\% CI $[-0.058, -0.022]$, $p_\text{boot}<10^{-4}$, $p_\text{wilcox}=0.011$, Cohen's $d=-0.33$ (small effect). Together, these results indicate that SCAR's chunk savings are a large, robust effect, while the recall cost is measurable but small.

\textbf{Cross-Embedding-Model Transfer.} The same default $(\lambda=0.1, \gamma=0.80)$ transfers without recalibration to two additional embedding models with different dimensionalities and similarity distributions: \texttt{BGE-large-en-v1.5} (12.7\% chunk reduction at 0.913 recall) and \texttt{zembed-1} (17.2\% chunk reduction at 0.956 recall), confirming that the relative-threshold formulation transfers across similarity scales without recalibration.

\subsection{Hyperparameter Robustness}
\label{sec:hyper}

Sweeping $\lambda \in [0, 0.3]$ at $\gamma=0.80$ yields a smooth, monotonic trade-off (mean recall drops only 1.7 points over $[0, 0.20]$); $\lambda=0.1$ sits near the efficiency elbow, with any value in $[0.05, 0.15]$ comparable. A leave-one-corpus-out $(\lambda, \gamma)$ grid search places the default $(\lambda=0.1, \gamma=0.80)$ in a broad plateau where mean recall is $\geq 0.90$; CV-selected parameters trade chunks against recall inconsistently across held-out corpora, confirming the fixed default does not exploit per-corpus peculiarities.

\subsection{Robustness to Gold-Label Noise}
\label{sec:goldnoise}

To rule out the possibility that SCAR's advantage stems from LLM-generated gold favoring its heuristic, we re-evaluated every method under four alternative gold definitions derived from the human annotations (replace, permissive union, conservative intersection, and 2-of-3 majority on the triple-overlap subset). Across all variants, SCAR's recall lead over the Top-$k$ baseline on boundary-fragmented queries remains positive (gap range $+0.034$ to $+0.098$), and SCAR continues to match or approach Window$\pm 1$ recall while using fewer chunks. The annotation imperfections surfaced by IAA propagate symmetrically to all methods, so SCAR's measured advantage is not attributable to gold-label noise.

\subsection{Downstream Generation Quality}
\label{sec:ragas}

Retrieval recall is a proxy for downstream usefulness; ultimately what matters is whether a generator can answer correctly from SCAR's context. We evaluate on all 80 queries of the Microsoft 10-K corpus using GPT-4o-mini as the generator and an LLM-as-judge protocol following RAGAS \cite{es2025ragasautomatedevaluationretrieval} for faithfulness and answer relevancy.

Compared with Window ($\pm 1$), SCAR ($\pm 1$) produces essentially identical faithfulness (4.99/5 vs.\ 4.99/5), slightly higher answer relevancy (4.79 vs.\ 4.74), and a markedly smaller context: 934 tokens on average vs.\ 1{,}281, a 27.1\% token reduction. Context precision improves from 0.17 to 0.23 while context recall (fraction of gold chunks present) remains at 0.99. These results indicate that SCAR's leaner context does not degrade grounded-answer quality at generation time.

\section{Conclusion}

SCAR is a training-free, retrieval-time policy that selectively expands context by weighing query-neighbor relevance against a structural continuity penalty, with a relative threshold that transfers across embedding models without recalibration. Across 160 boundary-fragmented queries it cuts chunk volume by 22.9\% versus static windowing ($p<10^{-4}$, $d=-1.49$) at a 3.9-point recall cost, while RAGAS on the 10-K corpus confirms faithfulness is preserved at a 27.1\% smaller context, making it a drop-in efficiency upgrade for production RAG. Its scope is adjacent-chunk fragmentation: it cannot recover evidence the retriever fails to rank in top-$k$, nor non-adjacent evidence; learned continuity penalties and non-adjacent logical-neighbor expansion are natural next directions. A further limitation is that SCAR assumes a needed adjacent chunk retains enough standalone query relevance to pass the expansion gate: for semantically thin continuations (e.g., table values separated from their headers, or paraphrased restatements), the neighbor's low direct relevance can cause SCAR to skip evidence that static windowing would recover. Structure-aware chunk construction (contextual prepending, header/row carryover) mitigates this and is complementary to SCAR.

\section*{GenAI Usage Disclosure}

Generative-AI components are part of the retrieval pipeline under evaluation: \texttt{text-embedding-3-large}, \texttt{BGE-large-en-v1.5}, and \texttt{zembed-1} produce the dense chunk and query embeddings, and the 320-query evaluation benchmark was bootstrapped using an LLM-assisted query-generation pipeline, followed by expert manual verification and the three-rater inter-annotator-agreement study described in Section~\ref{sec:experimental_setup}. The downstream-generation evaluation in Section~\ref{sec:ragas} uses GPT-4o-mini as the generator under an LLM-as-judge (RAGAS) protocol.

Generative-AI assistants were additionally used to draft and revise prose passages and to suggest code refactors during analysis. All experimental design, hyperparameter choices, statistical methodology, dataset curation, annotation procedures, manual adjudication of inter-annotator disagreements, and final scientific claims were formulated and verified by the author.

\bibliographystyle{ACM-Reference-Format}
\bibliography{scar}

\end{document}